\documentclass[a4paper]{jpconf}
\usepackage{graphicx}

\begin{document}
\title{Modeling subgrid combustion processes in simulations of thermonuclear
supernovae}

\author{Dean M Townsley$^1$, Alan C Calder$^2$ and Broxton J Miles$^3$}

\address{$^1$ Department of Physics and Astronomy, University of Alabama, Tuscaloosa, AL 35487-0324, USA}

\address{$^2$ Department of Physics and Astronomy, Stony Brook University, Stony Brook, NY 11794-3800, USA}

\address{$^3$ Department of Physics, North Carolina State University, Raleigh, NC 27695-8202 USA}

\ead{Dean.M.Townsley@ua.edu}

\begin{abstract}
Supernovae of type Ia are thought to arise from the thermonuclear incineration of a carbon-oxygen white dwarf stellar remnant.
However, the detailed explosion scenario and stellar evolutionary origin scenario -- or scenarios -- which lead to observed supernovae are still quite uncertain.
One of the principal tests of proposed scenarios is comparison with the explosion products inferred, for example, from the spectrum of the supernovae.
Making this comparison requires computation of the combustion dynamics and products through simulation of proposed scenarios.
Here we discuss two specific proposed explosion scenarios, the deflagration-detonation transition and the helium shell double detonation,
With these two examples in mind, we proceed to discuss challenges to computational modeling of the combustion taking place in these explosions.
Both subsonically and supersonically propagating reaction fronts are discussed, called deflagrations and detonations respectively.
Several major stages of the combustion occur on length and time scales that are many orders of magnitude smaller than those accessible in simulations of the explosion.
Models which attempt to capture this sub-grid behavior and the verification of those models is briefly discussed.

\end{abstract}

\section{Introduction}

Type Ia supernovae have spectral properties, including a lack of hydrogen or helium and a strong silicon feature, that distinguish them as a well-characterized class.
Thermonuclear incineration of a white dwarf star can explain these astrophysical stellar transients, with radioactive decay of nickel providing the power in their optically bright phase.
Here we discuss two example scenarios from those proposed and use these to then motivate a brief discussion of the challenges met in simulating these scenarios due to the scales of the combustion fronts
involved.

\section{Examples of proposed scenarios}

While it is fairly clear that Type Ia supernovae are produced by the incineration of a carbon-oxygen rich white dwarf star, there are a variety of suggested scenarios for how this might occur.
(See \cite{SeitenzahlTownsley2017,RopkeSim2018} for recent reviews.)
Specific scenarios generally have both advantages and shortcomings, and each is at a different level of development in modeling of its detailed predictions.
Here we will briefly discuss just two scenarios, the deflagration-detonation transition (DDT) and the helium shell double detonation,
with the aim of motivating the need to perform accurate and verifiable simulations that can be used to test these predictions against observations.
Our choice of these scenarios follows our desired to explain normal SNe~Ia, there are other scenarios for less normal explosions, such as merger and interaction that could lead to a superluminous event
\cite{Noebaueretal2016} or deflagration-only \cite{Roepkeetal2007}.
Many aspects of these scenarios depend on the combustion mode taking place and these will be discussed more fully in section \ref{sec:combustion}.
Each combustion mode burns either helium or carbon and oxygen to heavier elements in a localized reaction front.
However while a deflagration front will move subsonically, a detonation will move supersonically, and this aspect will cause them to have very different impacts on the overall explosion.

\subsection{Deflagration-detonation transition}

White dwarfs have a convenient natural ignition mechanism due to the Chandrasekhar mass limit.
Near this limiting mass, around 1.4~M$_\odot$, due to the electrons which support the star becoming relativistic, the central density becomes a singular function of the star's mass.
Thus if mass is added to a white dwarf star, some nuclear process will eventually take place due to the compression near the center of the star.
In the case of a carbon-oxygen rich interior, the process that takes place first is carbon fusion, which releases energy, creating a runaway that eventually incinerates the star.
However, there is a basic problem with this scenario.
If a Chandrasekhar mass white dwarf is incinerated, it makes almost purely iron-group material, whereas observed type Ia supernovae have half of the mass of their ashes in silicon group material.

The requisite silicon-group material can be created if the combustion occurs slowly enough that the white dwarf can expand to lower densities before all of it burns.
The first broadly successful thermonuclear supernova model, the W7 case from \cite{NomotoThielemannYokoi1984}, did this somewhat artificially with a tunable parameter.
The deflagration-detonation transition model \cite{Khokhlov1991} proposes that the burning begins slowly in the subsonic deflagration mode, and then transitions to the faster detonation mode.
This allows the star to expand enough to give ejected yields similar to those observed in actual objects \cite{HoeflichKhokhlov1996}.

However there are many assumptions that must be made for this scenario to work, and several have turned out to be genuinely challenging to make realistic.
There are many issues, but we will highlight just a few.
It is not easy to construct stellar systems in which the white dwarf will increase in mass in the necessary way.
Also, the distribution of ignition points of the deflagration stage needs to be fairly symmetric, but recent simulations of this process find that it should not be so \cite{Zingaleetal2011,Nonakaetal2012}.
Additionally, while 1-dimensional models provided a good match to the variety of explosions observed,
recent simulations of multi-dimensional models indicate that the relation between brightness and decline rate is opposite that observed \cite{Seitenzahletal2013,Milesetal2016}.
Thus while something like the deflagration-detonation transition model has been favored for some time, and remains a leading scenario, there are reasons to investigate others as well.

\subsection{Helium shell double detonation}

As noted above, Chandreskhar mass white dwarfs, if incinerated, will make more iron-group and less silicon-group elements than observed in type Ia supernovae.
Lower mass white dwarfs, when incinerated quickly by, for example, a detonation, will make a balance of element more similar to that observed \cite{Shenetal2018}.
However, a viable ignition mechanism for sub-Chandrasekhar mass white dwarfs is more elusive and not as natural as in the Chandrasekhar-mass case.
One possible ignition scenario is called the helium shell double detonation.
In this scenario, burning starts in a helium layer, which only needs to be a few hundredths of the star's mass to ignite.
After a detonation sweeps through the helium shell, a shock is left propagating inward through the carbon-oxygen core, which eventually focuses enough to ignite detonation of the rest of the star.
When first proposed, this scenario was thought unlikely because a shell thick enough to host a detonation would make a titanium and nickel-rich outer layer that is not observed in supernovae
\cite{LivneArnett1995}.

More recently, as the deflagration-detonation transition scenario has run into challenges, the double detonation scenario has been revisited.
It was found that if the helium layer burning turned out to be less complete than previously predicted, this would make a viable scenario \cite{Kromeretal2010}.
However, it appeared that very strong enrichment of the helium layer by underlying carbon would be needed to make a layer that would host a detonation while not producing too much titanium and chromium.
Since that work it has been found that by including more complete reactions in the simulation and a critical nitrogen isotope in the helium layer, this may be a workable scenario \cite{ShenMoore2014}.
It is also promising that simulations of the early stages of the merger process of two white dwarfs show that ignition of the helium layer may be a generic feature of white dwarf mergers \cite{Guillochonetal2010}.

\section{Modeling small-scale reaction structure in burning}
\label{sec:combustion}

Challenges in accurately simulating thermonuclear supernovae arise largely from the wide range of scales that the combustion and explosion processes cover.
To simulate the explosion, the whole star, $~10^8$~cm, must be simulated.
However, combustion processes occur on scales as small as a micron.
Here we will discuss briefly how these challenges are met for the two major modes of combustion active in thermonuclear supernovae: deflagration and detonation.

\subsection{Deflagration}

We use the term deflagration to indicate reaction fronts that propagate due to thermal diffusion.
In thermonuclear supernovae these fronts typically burn carbon and oxygen to silicon or iron group elements.
Their speed is determined by the balance of heat diffusion and reaction and has been computed in both steady state and hydrodynamic calculations
\cite{TimmesWoosley1992,ChamulakBrownTimmes2007}
and demonstrated to be stable \cite{GlazyrinBlinnikovDolgov2013}.
In quiet, i.e.\ laminar, flow, this flame has a 1-dimensional structure that proceeds from unburned to burned across a multi-stage reaction front.
This is typically thought of as a flame surface when the reaction stages are short compared to the lateral extent of the flow geometry.

\begin{figure}[h]
\begin{minipage}{2in}
\includegraphics[width=\textwidth]{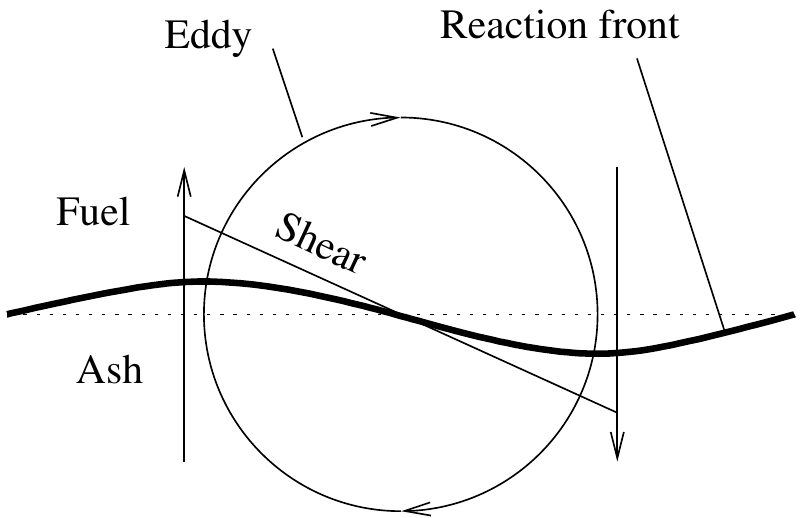}
\caption{
\label{fig:shear_diag}
Action of an eddy in the flow to shear and wrinkle the reaction front surface.
}
\end{minipage}\hspace{2em}%
\begin{minipage}{4in}
\includegraphics[width=\textwidth]{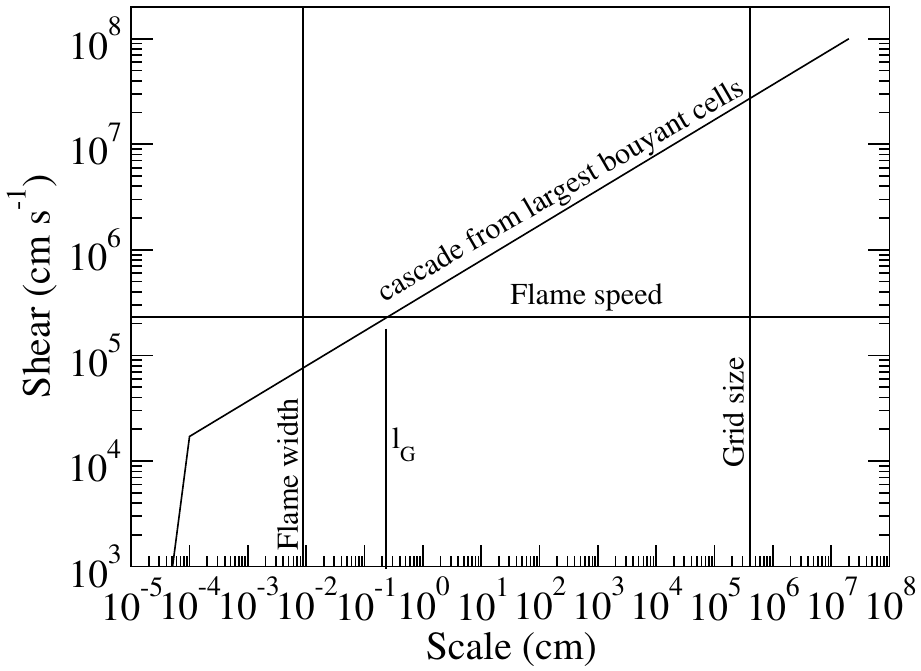}
\caption{
\label{fig:shear_plot}
Comparison of shear strength in turbulent cascade to velocity and length scales of the flame.
(adapted from \cite{JacksonTownsleyCalder2014})
}
\end{minipage}
\end{figure}

In the presence of shear flow the flame surface will become wrinkled as indicated in figure \ref{fig:shear_diag}.
Turbulent shear is the type of shearing flow of primary interest at subgrid scales in thermonuclear supernovae.
A distinctive feature of turbulence is that shearing occurs on all scales in a mostly random way, though the shearing strength depends on scale in a predictable way.
Figure \ref{fig:shear_plot} compares the strength of the shearing field due to Kolmogorov turbulence to the properties of the flame approximately midway through a thermonuclear supernova.
Several important scales are labeled.
The grid size at which typical simulations are performed is a few $10^5$~cm.
The width of the carbon burning stage of the laminar flame is many orders of magnitude smaller than this, of order $10^{-2}$~cm.
Each of these scales is indicated with a vertical line.

The spectrum of turbulent shear on many scales will generally act to wrinkle the flame surface.
This action, however, does depend on a competition between the flame propagation and the eddy turnover.
Considering the flow shown in Figure \ref{fig:shear_diag}, it is clear that if the flame surface propagates faster than the eddy overturns, the eddy will not effectively wrinkle the flame.
Additionally, wrinkled flames tend to self-smooth by burning out small-scale structure.
As a result of this action, there is a scale below which the flame surface will not be effectively wrinkled, termed the Gibson scale, and indicated by $l_G$ in Figure \ref{fig:shear_plot}.
This is the scale below which the flame speed is faster than the typical shear velocity of a turbulent eddy.

Capturing and modeling this structure of subgrid turbulence flame interaction is done with a model, which inherently has some free parameters that must be chosen
\cite{JacksonTownsleyCalder2014,Schmidtetal2006}.
The typical strategy is to treat some coarsened structure at the grid scale, which represents the coarse propagation of the reaction front.
The speed at which this coarsened structure propagates and other aspects of its behavior are then controlled with a model.
This model will then include various approximations and features intended to capture the physics of the subgrid turbulence-flame interaction.
The overall effect is much like accelerating the flame propagation, leading to a turbulent flame speed $s_t = \Xi s_\ell$,
where $s_\ell$ is the laminar flame speed and $\Xi$ is a wrinkling factor enhancing the propagation.

\subsection{Detonation}

\begin{figure}[h]
\begin{minipage}[b]{4in}
\includegraphics[width=\textwidth]{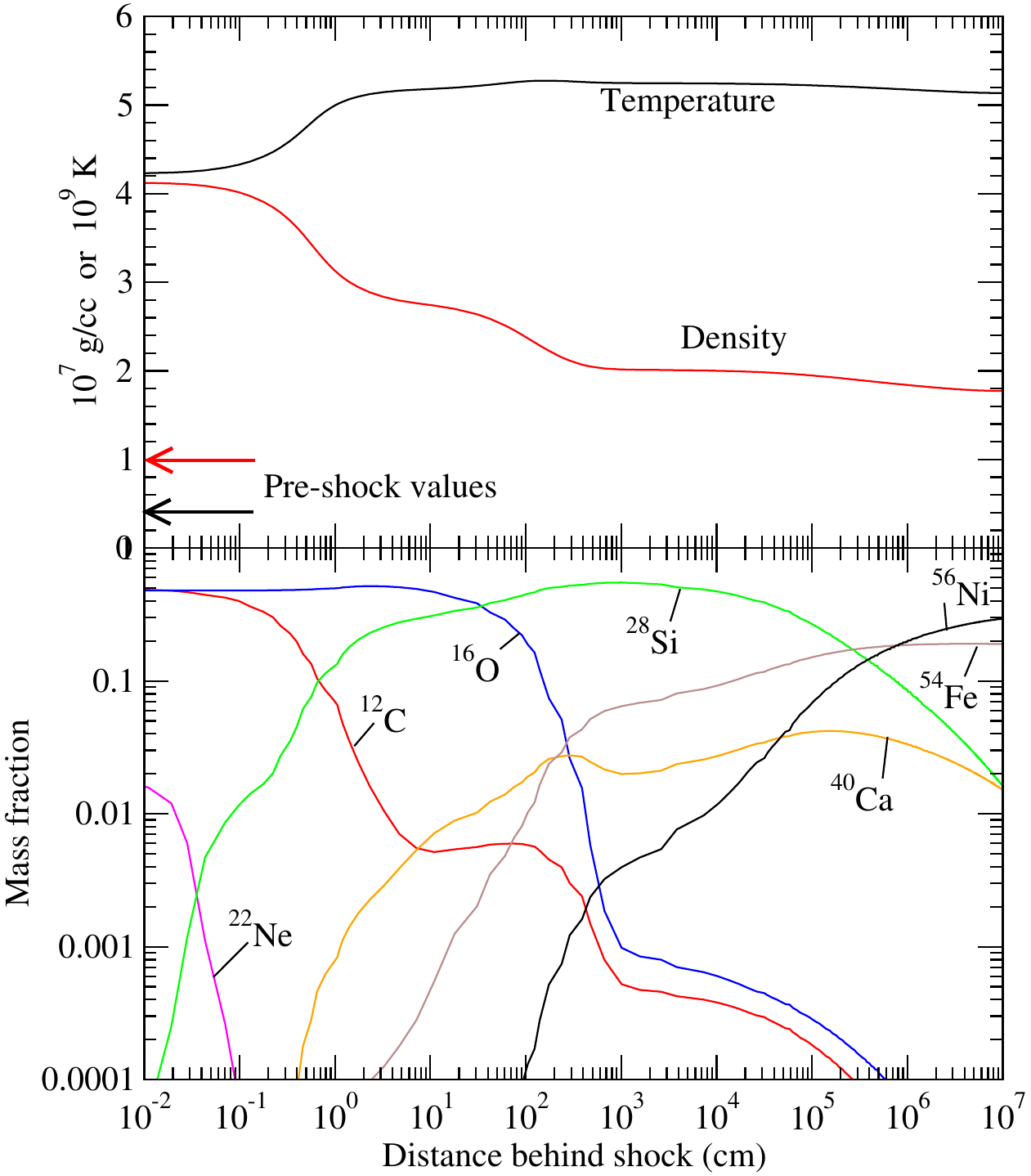}
\end{minipage}\vspace{2em}%
\begin{minipage}[b]{2in}
\caption{
\label{fig:det}
Thermodynamic and composition structure behind the shockfront in a detonation in a carbon-oxygen mixture at a density of $10^7$~g~cm$^{-3}$.
}
\end{minipage}
\end{figure}

Detonations are reaction fronts that are propagated through fuel by a shock that is self-sustained by the energy released in burning.
In thermonuclear supernovae, detonations also present the challenge of unresolved detailed reaction kinetics in simulations that compute full-star explosions.
Figure \ref{fig:det} shows the basic thermodynamic and composition structure of a detonation propagating in a carbon-oxygen mixture.
See \cite{Townsleyetal2016} for more details on how this steady-state structure is computed.
The various thermodynamic properties -- temperature and density -- and the mass fraction abundance of the most abundant species are shown as a function of distance behind the propagating shock.
This figure shows the detonation structure for a moderate density expected in the exploding white dwarf star, $10^7$~g~cm$^{-3}$.

As mentioned previously, the typical grid scale for a full-star simulation is a few $10^5$~cm.
This means that at a wide range of densities, the length scale for consumption of the fuel, but carbon and oxygen, is unresolved.
Exactly which stages are unresolved and how complete the burning can be depends strongly on the density of the fuel.
Also the overall size of the star, around $10^8$~cm, is important for determining when the burning stops.
At densities above approximately that shown in Figure \ref{fig:det}, $10^7$~g~cm$^{-3}$, there is enough time and space for the burning to produce iron-group material,
as shown near the large scales on the right side of Figure \ref{fig:det}.
At lower densities, more stages become resolved and the burning will be truncated with the production of silicon-group material.

One of the constraints imposed by the overall size of the star is that the detonation front will typically be curved instead of planar \cite{DunkleySharpeFalle2013}.
This arises just because the detonation must originate in some localized place and then move radially away from that spot.
The resulting propagation will, in general, not align with the spherical structure of the star due to the initiation point being away from the star's center.
While this curvature is macroscopic, that is, resolved in the simulation, the resulting weakening of the shock and post-shock flow results in slower burning even on microscopic scales.
This leads to early truncation of burning and the need to treat these unresolved scales with some model \cite{Milesetal2018}.

\subsection{Method and verification}

The small-scale physics governing combustion in thermonuclear supernovae must be accounted for in two distinct ways.
First, explosion simulations, which are coarse in resolution by necessity of capturing the whole star, must use some model for the combustion.
These models have been discussed briefly above.
For deflagrations, techniques include flame thickening \cite{Calderetal2007,Townsleyetal2009,Townsleyetal2016} or front-tracking \cite{RopkeNiemeyerHillebrandt2003}.
For detonations, generally some form of reaction limiting is used \cite{Kushniretal2013,Townsleyetal2016,Shenetal2018}.
In either case, frequently a reduced nuclear reaction network is used in the simulation as well.

The second component of the calculation begins by tracing the history of fluid elements in the simulated explosion.
This produces density and temperature histories, $\rho(t)$ and $T(t)$ respectively, each of which we call a track.
In post-processing, a complete nuclear reaction network, usually with 200 or more species, is used to compute, for each track, the ashes of the overall combustion process.
These yields can then be compared to observations, either via the spectrum of the supernova, or via abundances of material in the solar system.

Our recent work has focused on using ancillary information from the simulation, beyond the $T$, $\rho$ history, to make the post-processed abundances more physically realistic and verifiable.
This is necessitated by the fact that, as shown in previous sections, many combustion stages are unresolved, and therefore are not represented accurately in the track produced by the simulation.
In order to compensate for this, we use the physics of the particular combustion mode to reconstruct these unresolved combustion stages.
In the case of the deflagration, the pressure near the reaction front is used to inform a self-heating calculation that mimics how a laminar flame behaves \cite{Townsleyetal2016}.
For detonations, we use information about the curvature and density gradients to reconstruct the unresolved stages based on steady-state models \cite{Milesetal2018}.
Verification now becomes more achievable, as the detailed model of the unresolved combustion stages can be closely compared to the outcome of the simulation.
Also, the uncertainties are controlled by the degree to which approximations like the steady state and constant pressure assumptions are satisfied, which can be mostly quantified.

\section{Summary and conclusion}

Simulations of proposed scenarios for thermonuclear supernovae must accurately and verifiably capture the physics of combustion processes that occur at scales smaller than the grid scale.
Here we have briefly discussed the combustion modes involved, techniques used to model them, and some example scenarios in which these combustion modes appear.
Modelling the unresolved combustion is addressed in two steps.
Hydrodynamic simulations of the full explosion use a model which is intended to capture the energy release dynamics of the burning.
The thermal history determined by these simulations are then used along with more detailed reaction physics and steady-state models for unresolved processes to compute final yields from the explosion.
Comparison of these yields to those inferred from observed supernovae will then allow identification of which scenarios might occur in nature, how often, and how they might be distinguished.

\ack

DMT and BJM received support from NASA through the Astrophysics Theory Program (NNX17AG28G).
BJM was also supported in part by the United States Department of Energy under an Early Career Award (DOE grant no. SC111026) and by the Research Corporation for Science Advancement under a Cottrell Scholar
Award to Carla Fr\"olich.
ACC received support from the US Department of energy under grant DE-FG02-87ER40317.

\section*{References}
\bibliography{townastronum}

\end{document}